\documentclass[%
 reprint,
 amsmath,amssymb,
 aps,
]{revtex4-2}

\usepackage{graphicx}% Include figure files
\usepackage{dcolumn}% Align table columns on decimal point
\usepackage{bm}% bold math
\usepackage{xcolor}
\usepackage{hyperref}% add hypertext capabilities
\begin{document}

\preprint{APS/123-QED}

\title{Characterizing seismic isolation using convolutional neural networks and Wiener filters}% Force line breaks with \\
%\thanks{A footnote to the article title}%

\author{Artem Basalaev}
 \email{artem.basalaev@physik.uni-hamburg.de}

 \author{Jan-Niklas Feldhusen}
 \email{jan-niklas.feldhusen@physik.uni-hamburg.de}

\author{Oliver Gerberding}%
 \email{oliver.gerberding@physik.uni-hamburg.de}
 
% \altaffiliation[Also at ]{Physics Department, XYZ University.}%Lines break automatically or can be forced with \\
\affiliation{%
 Institute of Experimental Physics, University of Hamburg,\\ Mittelweg 177, 20148 Hamburg, Germany
}%

%\collaboration{MUSO Collaboration}%\noaffiliation

\date{\today}% It is always \today, today,
             %  but any date may be explicitly specified

\begin{abstract}
We investigate seismic motion propagation through a passively isolated mechanical system, using Wiener filters and convolutional neural networks with time-dilation layers. The goal of this study was to explore the capabilities of neural networks and Wiener filters in characterizing a mechanical system from the measurements. The mechanical system used is a testbed facility for technology development for current and future gravitational wave detectors, ``VATIGrav'', currently being commissioned at University of Hamburg. It consists of a large vacuum chamber mounted on four active vibration isolators with an optical table inside, mounted on four passive vibration isolators. In this paper we have used seismic data recorded on the ground and on the optical table inside the chamber. The data were divided in 6 hours for training and another 6 hours for validation, focusing on inferring 150-second stretches of time series of table motion from the ground motion in the frequency range from 0.1\,Hz to about 50\,Hz.
We compare the performance of a neural network with FTT-based loss function and with Huber loss function to single-input, single-output (SISO) and multiple-input, single-output (MISO) Wiener filters. To be able to compute very large MISO Wiener filters (with 15,000 taps) we have optimized the calculations exploiting block-Toeplitz structure of the matrix in Wiener-Hopf equations. We find that for the given task SISO Wiener filters outperform MISO Wiener filters, mostly due to low coherence between different motion axes. Neural network trained with Huber loss performs slightly worse than Wiener filters. Neural network with FFT-based loss outperforms Wiener filters in some frequency regions, particularly with low amplitudes and reduced coherence, while it tends to slightly underestimate the peaks, where Wiener filters perform better.

\end{abstract}

%\keywords{Suggested keywords}%Use showkeys class option if keyword
                              %display desired
\maketitle

%\tableofcontents
\section{\label{sec:Introdution}Introduction}

Seismic isolation plays a crucial role in enabling design performance of current and next generation ground-based gravitational wave detectors, such as Advanced LIGO \cite{LIGOScientific}, Advanced Virgo~\cite{VIRGO}, KAGRA~\cite{KAGRA}, GEO600~\cite{Luck:2010} and the future Einstein Telescope~\cite{ET} and Cosmic Explorer~\cite{CosmicExplorer, CosmicExplorer2}. Natural and human-made (e.g. by industrial equipment, trucks) seismic vibrations in the lower audio-frequency band affect gravitational wave detectors in two main ways, depending on their coupling mechanism.

Vibrations can couple directly into the detection band, which for current LIGO and VIRGO detectors starts at~10 Hz~\cite{LIGOScientific} and for future detectors such as Einstein Telscope are expected to be extended down to a few~Hz~\cite{ET_Sensitivity}. These vibrations directly affect the signal to noise ratio and related metrics such as the detector's astrophysical reach in Megaparsecs.
    
Vibrations in and below the detection band may introduce large amplitude motions that interferometer (IFO) control systems cannot fully compensate, resulting in the optics drifting out of alignment and lock loss. Even if the lock loss does not occur, the required large amplitude actuation couples in the associated controls noise. This type of vibrations thereby indirectly limits sensitivity and can also reduce detector's duty cycle.

Seismic isolation systems, together with the suspension of mirrors, act as a filter, preventing vibrations at frequencies in the detection band from propagating to the optics. Motion at frequencies below the detection band contributes to the overall motion of the optics which determines the required control authority~\cite{vandongen2023}. It can also couple into the measurement by non-linear effects, such as scattered light~\cite{Canuel2013}. Especially at the suspensions resonances this motion is critical, due to the strong motion amplification. 

This can be counteracted by using active seismic pre-isolation platforms onto which the suspension chain is mounted, such as Internal Seismic Isolation (ISI) platform at LIGO~\cite{Matichard2015} and similar solutions at gravitational wave detector prototypes \cite{DiPace2022,AEI10mprototype}. This approach has some challenges, notably it is limited by the self noise of the sensors used in pre-isolation control loops. Sensor self noise is propagating through the suspension chain as spurious motion, affecting measurement degrees of freedom of the detector and limiting its overall sensitivity. More sensitive local, inertial sensors and rotation sensors are being developed to improve this \cite{Carter2024, WashingtonRotationSensors, 6DPaper}.

At University of Hamburg, we are setting up a test environment to study these effects on a smaller scale, consisting of a vacuum chamber with active pre-isolation, called \textit{Va}cuum system for \textit{t}h\textit{i}rd generation \textit{grav}itational wave detector prototyping, or VATIGrav. It is intended for testing and the development of local and inertial sensors mounted on suspensions. Especially the compact interferometric sensors, which have gained strong interest in recent years~\cite{HoQiSensitivity,vandongen2023}, because they can provide greatly improved noise performance in comparison to the currently used shadow sensors \cite{BOSEM} at LIGO. We are developing one concept of such a sensor in-house, based on the technique of deep-frequency modulation~\cite{gerberding2021}, \textit{Co}mpact \textit{B}alanced \textit{R}eadout \textit{I}nterferometer sensor -- COBRI.

In this paper we analyze seismic motion propagating through the mechanical system of passive seismic isolation of VATIGrav, and attempt to  infer the motion of the optical table inside the vacuum chamber from the measurement of ground motion, using convolutional neural networks (CNNs) and Wiener filters. This work is focused on exploring capabilities of Wiener filters and CNNs, but will also contribute towards future improvements of the seismic isolation of VATIGrav, involving more sensors and better active control. 

The Wiener filter and neural network approach developed here could be of general interest for seismic isolation in gravitational wave detectors. We utilize 3-axis data for both input and output $(X_\text{gnd}, Y_\text{gnd}, Z_\text{gnd}) \rightarrow (X_\text{table}, Y_\text{table}, Z_\text{table})$ and comparatively long time series to capture a broad frequency range. 

Previous works featuring neural networks for time series include~\cite{pendulum_machine_learning}, where deflection angles of a seismically isolated pendulum were predicted. Another relevant study is the prediction of terrestial gravity fluctuations (Newtonian) noise from an array of seismic data~\cite{BertramNN}, which motivated the overall design of our CNN. Wiener filters were also studied for suppression of Newtonian noise~\cite{harms} as well as for active vibration isolation at 40 m prototype interferometer lab at Caltech~\cite{wf_paper} and LIGO detectors~\cite{DeRosa_2012}.

\section{\label{sec:vatigrav}VATIGrav: testbed for next generation GW detector technology development}

\begin{figure}[b]
\includegraphics[scale=0.45]{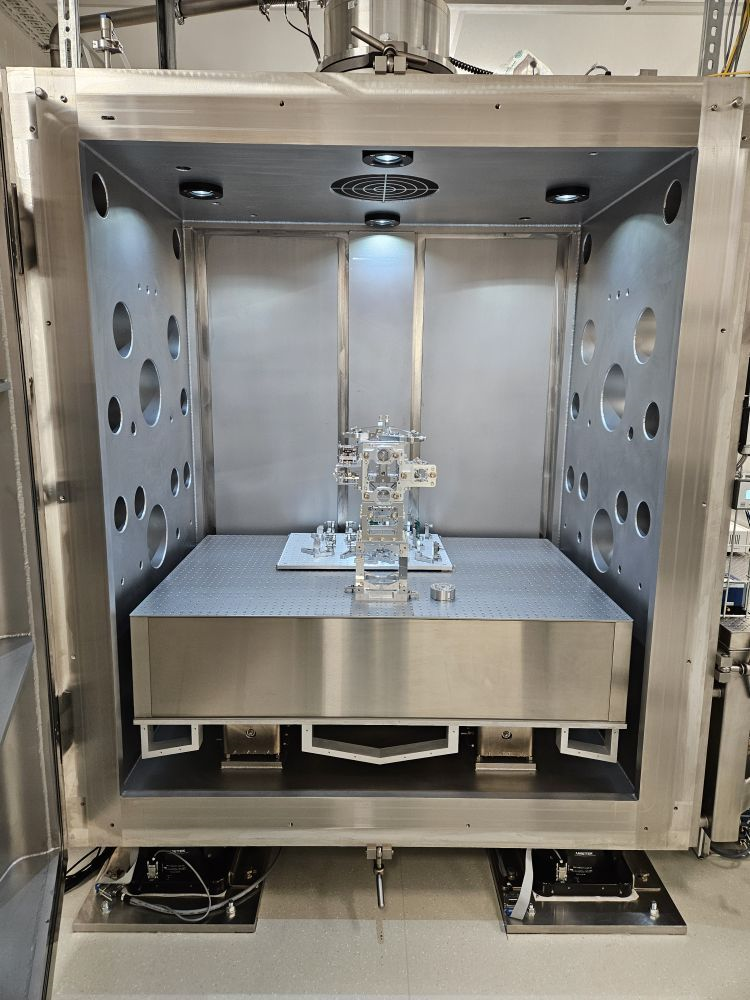}% Here is how to import EPS art
\caption{\label{fig:vati} VATIGrav chamber. The turbopump can be seen on top of the chamber. The HRTS suspension is mounted on the optical table inside. Below the optical table the aluminum weight holder as well as CM-1 isolators can be seen. The chamber is standing on STACIS III isolation feet.  }
\end{figure}

VATIGrav, shown in Fig.~\ref{fig:vati}, is a testbed facility for technology development for current and future gravitational wave detectors, which determines its design. It is a cubic chamber approx. 1.5 m long, 2 m wide and 2.5 m high, weighing approximately 5.5 tonnes. The chamber features double doors in the back and a single door in the front, resulting in internal dimensions of $1.02 \times 1.74 \times 1.51~\text{m}$. That volume provides sufficient space for mounting suspension chains and assembling interferometers needed for testing of the local sensors, with easy accessibility from either side to adjust the setup between test runs. The vacuum system features a turbopump 
 mounted on top and two pre-pumps located outside of the laboratory to reduce noise. The vacuum system is designed to achieve high vacuum levels (so far the lowest pressure reached is $1.9 \cdot 10^{-6} \text{mBa}$ after a few days of pumping).

The optical table inside VATIGrav (240 kg) is placed on four Minus-K 250~CM-1 passive vibration isolators (12 kg each). CM-1 isolators perform best under weight close to maximum payload capacity, which in our setup means up to 190 kg of additional weight. The isolators can be tuned with the help of vertical stiffness adjustment screw (coarse) and load adjustment crank (fine). To be able to operate under optimal weight while changing experimental setups, we designed an underside weight holder where stainless steel rods can be inserted for proper balancing (see Fig.~\ref{fig:vati}). Besides, smaller stainless steel cylinders can be placed on the optical table. 

The chamber itself (weighing approximately 6 tonnes with payload) is placed on four Ametek/TMC STACIS III active vibration isolators. The latter are equipped with three orthogonally oriented ``GS-ONE-LF'' geophones and actuators each. Based on the readings from geophones and preset controller, each of the units independently compensates the vibrations it is sensing via a feedback loop. Independent operation ensures that the units are agnostic to their relative placement under the chamber, but introduces a penalty since any tilts of the chamber cannot be accounted for and compensated. The controlling hardware and software is commercial and closed-source without a possibility to make any adjustments; the only possible adjustment is a limited tuning of the control transfer function of the unit, that was performed by the TMC personnel after installation of the chamber.

\subsection{\label{sec:vatigrav_performance}Seismic isolation of VATIGrav}

We measured the seismic isolation performance of VATIGrav with two seismometers, a Trillium 360, which is placed on the center of the optical table in the chamber and an STS-2.5 on the ground nearby. Data recorded at 100~Hz sampling rate at night were used in this study. It should be noted that the passive isolation is not expected to be fully optimal yet because the additional weight was not added for this measurement (resulting in the payload being too light) and CM-1 isolators were not properly tuned.

\begin{figure}[b]
\includegraphics[scale=0.45]{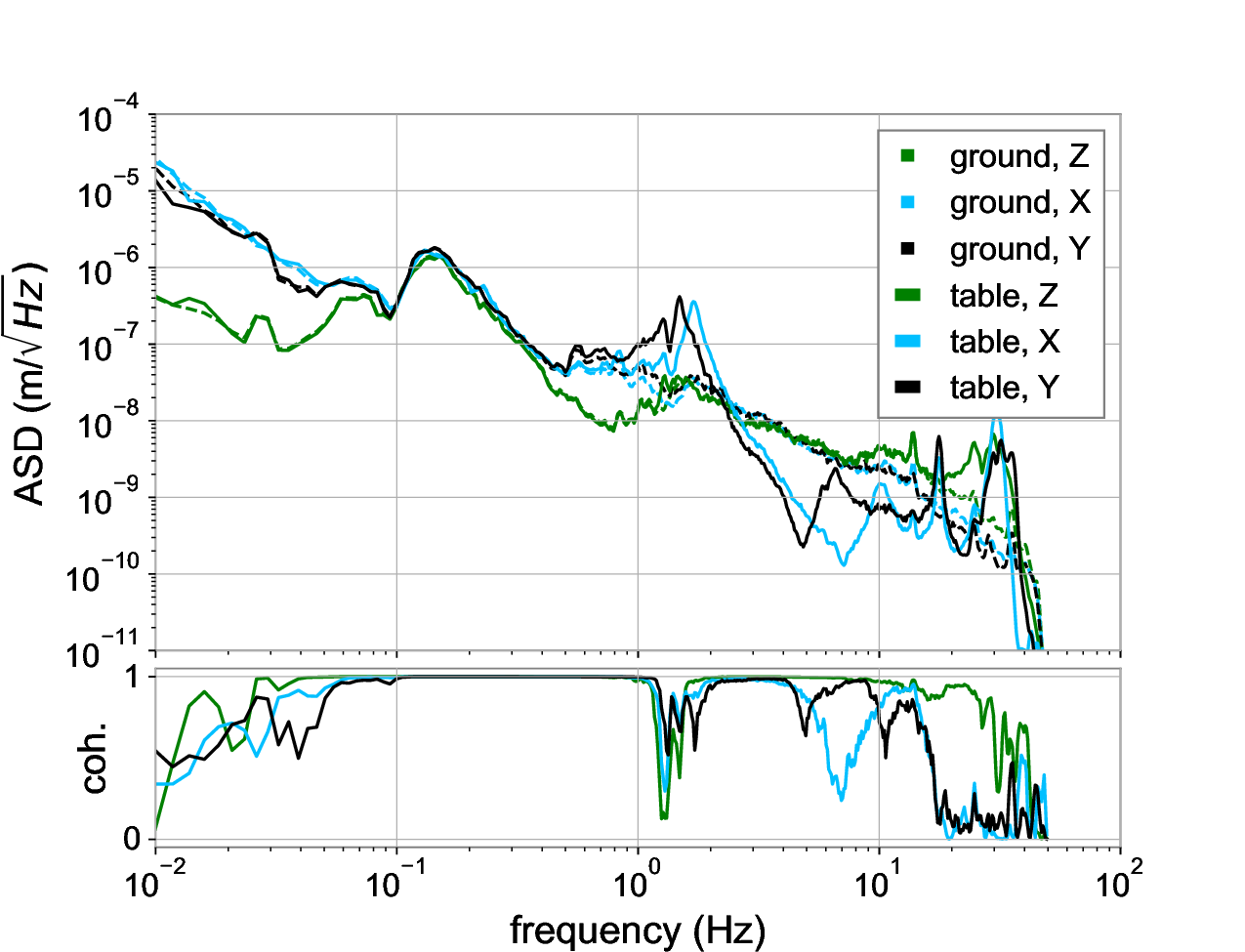}% Here is how to import EPS art
\caption{\label{fig:passive_iso_1h} Seismic measurements on the ground and on the optical table inside VATIGrav, STACIS III active isolation system \textit{is not} engaged.  Solid lines correspond to motion of the table, and dashed lines to motion of the ground. The lower subplot shows linear coherence between the table and the ground for same directions on the ground and on the table.}
\end{figure}
\begin{figure}[b]
\includegraphics[scale=0.45]{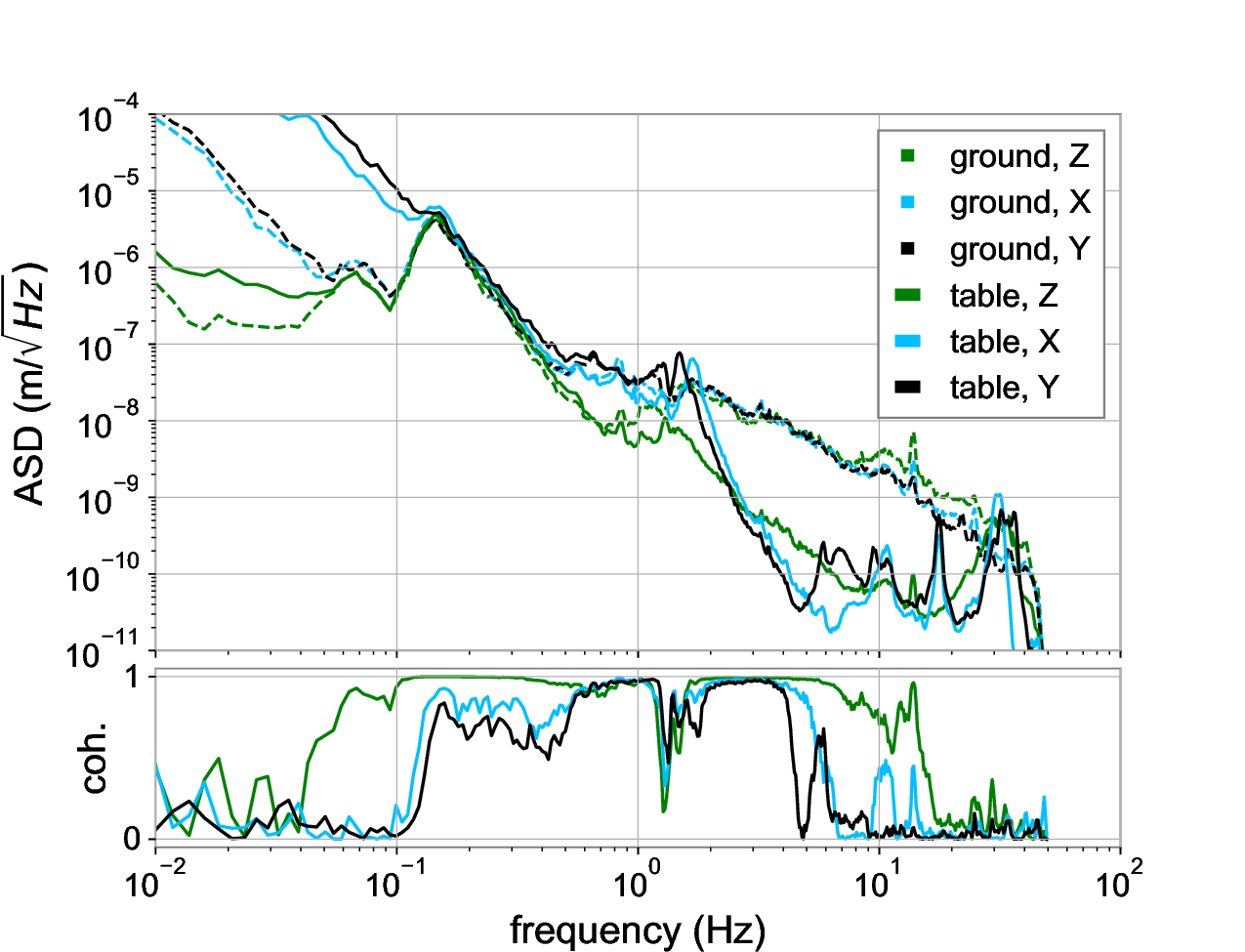}% Here is how to import EPS art
\caption{\label{fig:active_iso_1h} Seismic measurements on the ground and on the optical table inside VATIGrav, STACIS III active isolation system \textit{is} engaged. Solid lines correspond to motion of the table, and dashed lines to motion of the ground. The lower subplot shows linear coherence between the table and the ground for same directions on the ground and on the table.  }
\end{figure}

The amplitude spectral density (ASD) of the active isolation (STACIS III system turned on) is shown in Fig.~\ref{fig:active_iso_1h} and passive isolation (STACIS III system is off) performance is shown in Fig.~\ref{fig:passive_iso_1h}, for 1 hour of quiet data on two different nights. Since the measurements were taken in different time periods, background seismic levels are not the same, which is visible by the difference at the microseism peak at ~0.15 Hz -- higher for active isolation measurement. This however does not affect main conclusions about performance. 

On the plots with passive-only isolation, resonances are clearly visible for horizontal ($X$, $Y$) directions, at around 1.2-2 Hz. These resonances can be attributed to the under-loaded CM-1 isolators, which have a 0.5 Hz resonance frequency under optimal load~\cite{minusk}. No prominent resonance appears for vertical ($Z$) direction, and in general passive isolation in vertical direction does not appear to have an effect. This, as later discovered, was due to CM-1 isolators not set up properly, essentially resting on supports and not ``floating'' for the vertical motion. 

At frequencies above the horizontal resonances, passive isolation attenuates motion by close to 2 orders of magnitude at maximum for horizontal directions. At high frequencies some more resonances on the table appear, resulting in excess motion. Resonances around 18 Hz are the natural passive resonance of the STACIS III feet. The origin of other higher-frequency resonances is not so clear, they were not seen in a separate measurement on top of the chamber. That means that these resonances are occurring either in CM-1 isolators, or are related to the optical table and its payload. Their origin might become more clear when the passive isolation is optimized.

The active isolation system does a good job of suppressing these high-frequency resonances as well as 1.2-2 Hz resonances, and shows broadband improvement above ~0.2 Hz. At the same time, it is clearly visible that the active system injects motion below the microseism peak at ~0.15 Hz. This is not surprising given the noise introduced by geophones, their readout and not optimal independent control of each unit, which does not account for tilt.

Potential improvements of seismic isolation are possible for example by replacing the individual controllers with global control and filtering. Later seismometers (as higher precision sensors at low frequencies) could be  placed on the ground for feedforward- and potentially seismometers and geophones on the table in vacuum can be added for additional feedback control.

In the following chapters of this paper we explore whether we can infer the motion of the table, especially at low frequencies, from motion on the ground. We compare time series directly, in a random window a few seconds long, as well as amplitude spectral density (ASD) over one hour of data. A model transforming time series, that successfully captures motion characteristics on the table solely from the measurements on the ground, can later aid in a feedforward control scheme. However, the design and the implementation of such a scheme requires further work and is not discussed here.% is to be described in a follow-up paper.  

\section{\label{sec:nn}Neural Network design}

The network architecture we used to estimate the motion on the table from ground motion is an 1D-convolutional neural network (1D-CNN) with increasing time-dilation. We have chosen the CNN architecture over, for instance, simple dense networks (where all neurons are interconnected) because of its robustness against exploding or vanishing gradients in training due to regularized weights and fewer connections. The CNNs are therefore more efficient in training, requiring fewer computing resources and memory, which allows for more complex network structures with better potential for learning with a given amount of resources, especially for images, audio and video data~\cite{LeCun2015}. Seismic data is similar in its structure to audio data shifted to non-audible frequencies.

The receptive field of a CNN is fixed by its kernel size. To be able to infer low frequency motion, we need to have unbroken stretches of time series as long as possible. That would require increasing the number of layers or kernel size and would lead to more computational effort. We kept long stretches of time series (150~s, or 15,000 samples) but also used time-dilation to accommodate a large number of inputs with fewer computations. It applies the convolutional kernel over a lager field of data, and has been proven in audio-processing~\cite{Zhang2017} and in denoising of seismic data~\cite{Li2021}.

\begin{figure}[b]
\includegraphics[scale=0.5]{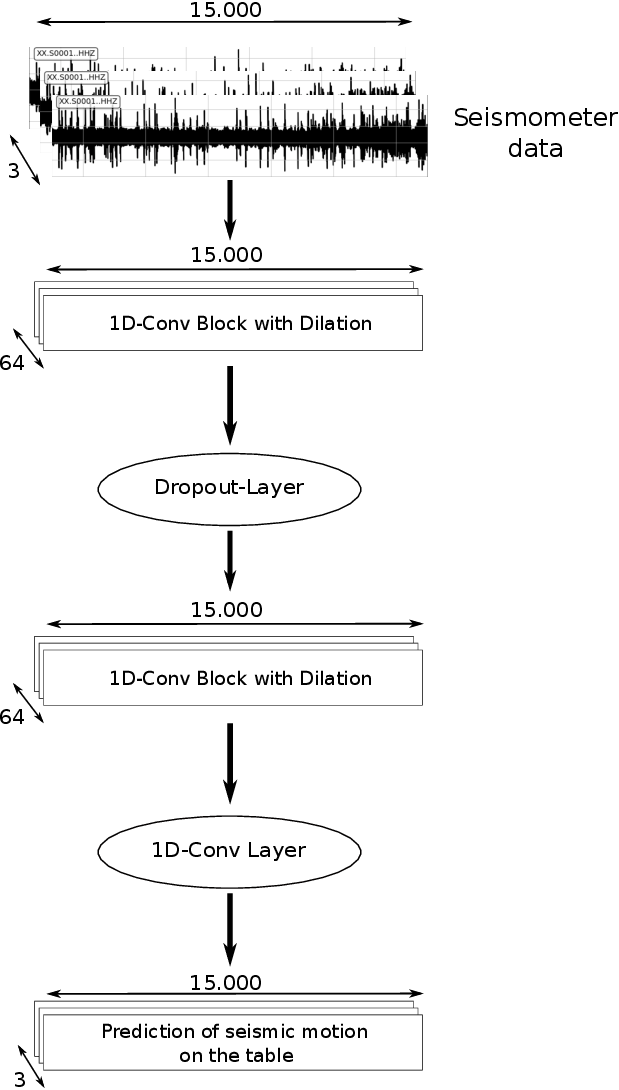}% Here is how to import EPS art
\caption{\label{fig:NN_structure} This sketch shows the structure of the convolutional neural network of this paper. It is a 1D convolutional neural network with dilation and consists of 2 convolution blocks, which contain six convolutional layers with increasing time-dilation. Between these blocks a dropout layer can be applied. At the end one 1D-convolutional layer is applied to reduce the dimensionality of data from 64 convolutional filters to 3 physical axes.}
\end{figure}

The architecture of the network can be seen in Fig.~\ref{fig:NN_structure}. It consists of two convolution blocks with a total of 13 1D-convolutional layers. One convolution block contains six layers with increasing time-dilation. In between the convolution blocks a dropout layer is applied, which randomly zeroes some outputs in the  training (with probability defined by dropout rate), to prevent overfitting. As an activation function the \texttt{leakyReLU} is used. The dimensionality is increased from three channels input data ($X$, $Y$, $Z$) to 64 convolutional filters. The last additional layer reduces the 64 filters back to three output channels. The number of time steps (15,000 samples) is kept constant throughout the network. 

For training the network focusing on ASD performance, a new loss function was defined, which the CNN aims to minimize:
\scriptsize
\begin{flalign}
    \text{loss} =1.25 \cdot 10^{-6}  \sum_{\text{bat., chan.}} \left(\sum_{\text{freq.}}\left|\frac{ln(\left| \text{rfft}_{\text{true}}\right|) - ln(\left|\text{rfft}_{\text{pred}} \right|)}{\text{fft sample frequencies}}\right| +\right. \nonumber \\
    + \left.\sum_{\text{freq.}}\left|\frac{arg( \text{rfft}_{\text{true}}) - arg(\text{rfft}_{\text{pred}})} {\text{fft sample frequencies}}\right|\vphantom{...}\right).\label{eq:lossFkt}
\end{flalign}
\normalsize

Instead of comparing the time series directly, the difference of the amplitudes and phases of fast Fourier transform (FFT) from the input and output data are calculated. (Here the \texttt{rfft} function used as input time series is real-valued.) The first component of loss is the amplitude difference, for which absolute values of the FFT at each frequency are used. A logarithm of these values taken first, with the result of the loss function being sensitive to logarithmic amplitudes scale, as used typically for ASD plots. The second component is phase difference, or difference in argument of the FFT at each frequency. Before summation over frequencies, both components are divided by the frequency. Since \texttt{rfft} frequencies are linearly spaced, this has an effect of reweighting to logarithmic scale. We consider the resulting log-log scale more natural (all spectra comparison plots in this paper feature log-log binning). The loss is computed for all channels and all batches and then summed up. It is then multiplied by a factor $1.25 \cdot 10^{-6}$, empirically chosen to bring a typical value close to 1.  Trained using this loss function, the network aims to get the correct amplitude of each frequency, by producing corresponding time series.

As an alternative, we have tried other loss functions that are applied directly to data arrays. Popular loss functions such as the mean squared error and the mean absolute error performed poorly for our case. Instead, we have used the Huber loss function, defined as:

\small
\begin{equation}
    \text{loss} =\sum_{\text{data samples}}{
    \begin{cases}
    \frac{1}{2}\left(\text{pred}-\text{true}\right)^2, \text{if}\ |\text{pred}-\text{true}| \leq \delta, \\
    \frac{1}{2}\delta \left( |\text{pred}-\text{true}| - \frac{1}{2} \delta \right)\ \text{otherwise}.\\
    \end{cases}
    } \label{eq:lossHuber}
\end{equation}
\normalsize

The Huber loss function is therefore quadratic for small deviations (less than $\delta$), and linear for larger deviations. This reduces sensitivity to outliers in the data~\cite{Huber}. In our case the optimal value for $\delta$ was found using a parameter sweep, explained below.

At each step in the training a subset of the data is used, a number of 150-second chunks defined by the batch size. Before the training, the 150-second chunks in the training dataset are randomly shuffled to reduce chances of overfitting.

The amount by which the network weights are allowed to change in each step of the training, or, equivalently, the step size along the gradient of the loss function, is defined by the learning rate (lr). This is an important hyper-parameter that balances reasonable speed of training (higher learning rate) with reducing the chance of ``overshooting'' the optimum (lower learning rate). 
It is advantageous to adjust the learning rate, starting with the higher value and decreasing it. A learning rate scheduler has been implemented, to decrease the learning rate for higher epochs:

\begin{equation}
    \text{lr} = 
    \begin{cases}
        \text{lr}_\text{start} & \text{for } \text{epoch} < 50,\\
        \text{lr}_\text{start}\cdot\frac{1}{\text{lr}_\text{decay} \cdot \text{epoch}} & \text{for } \text{epoch} \geq 50,\\
        \text{lr}_\text{min} & \text{if } \text{lr} < \text{lr}_\text{min}.
    \end{cases}
\end{equation}

The learning rate is computed for each epoch and depends on a start learning rate, the current epoch of the network, the learning rate decay and a minimal learning rate. As a minimal learning rate $\text{lr}_\text{min}$ = $10^{-7}$ was chosen. For the weight optimization the Nadam optimization algorithm is used, which takes as a parameter the weight decay to make the learning function smoother.

A separate validation dataset is created which is not used during training. We used this dataset to tune hyper-parameters of the network by using sweeps in wandb~\cite{wandb}.The hyper-parameters, for which the optimal values are found by the sweep, are batch size, $\text{lr}_\text{start}$, $\text{lr}_\text{decay}$, the dropout rate of the dropout layer, the weight decay, and the Huber $\delta$ when using the Huber loss function. A sweep searches in multidimensional space of these parameters, running training multiple times, each for 500 epochs, with the goal of minimizing loss on the validation dataset (validation loss). Bayesian hyper-parameter optimization is used, where each run's result guides subsequent values of hyperparameters to test. A minimum of 50 runs were used to optimize the network.

Finally, the run with best hyper-parameters performed again, this time with validation loss being the stopping criterion. We run the network training until the validation loss stops improving.

From the parameter sweep we got following result for the CNN with the FFT loss function:
\begin{itemize}
    \itemsep-0.5em 
    \item[-] batch size = 8,
    \item[-] $\text{lr}_\text{start}$ = $2.4236 \cdot 10^{-3}$,
    \item[-] $\text{lr}_\text{decay}$ = 0.3880,
    \item[-] dropout rate = 0.4,
    \item[-] weight decay = $1.5161 \cdot 10^{-5}$.
\end{itemize}

The optimal number of training epochs for the best run was 950, determined retrospectively after training for 1000 epochs.

And for the CNN with Huber loss function:
\begin{itemize}
    \itemsep-0.5em 
    \item[-] batch size = 4,
    \item[-] $\text{lr}_\text{start}$ = $2.4103 \cdot 10^{-3}$,
    \item[-] $\text{lr}_\text{decay}$ = 0.1789,
    \item[-] dropout rate = 0.1,
    \item[-] weight decay = $8.3082 \cdot 10^{-5}$,
    \item[-] Huber $\delta$ = 0.0659$\sigma$,
\end{itemize}

where $\sigma$ is the standard deviation of the data. The optimal number of training epochs for the best run was only 100, training longer actually increased the validation loss, i.e. overfitting occurred.

\section{\label{sec:wiener_filter}Wiener filter}

Wiener filters are widely used for correlated noise subtraction using references, and they are known to provide optimal subtraction of this kind in linear and stationary regime~\cite{harms}.

If there's a non-linear interaction, it is expected to be better captured by the neural network, and this is a reason why it is of interest to perform a comparison to Wiener filter to assess a potential neural network advantage. In particular, a Wiener filter is expected to perform worse for frequency ranges with low linear coherence. We calculate an estimate of linear coherence to aid the comparison.

In our case, the regime is not necessarily stationary. An adaptive Wiener filter could in principle account for this~\cite{wf_paper, DeRosa_2012}. 
Since we compare to a single neural network, we decided not to investigate adaptive Wiener filters here.
Instead, we compute 143 Wiener filters for the input data (corresponding to the number of 150-second intervals), and choose the best-performing Wiener filter among these, using a procedure described in~\ref{sec:wiener_filter_best}.

\subsection{\label{sec:wiener_filter_optimized}Optimized Wiener filter calculation}

A Wiener filter can be defined as follows~\cite{harms}:

$$\hat{y_n} = \sum_{k=0}^{N}{\mathbf{w_k}}\cdot \mathbf{x_{n-k}},$$

here $ \hat{y_n}$ is an estimation of the target signal at the time sample $ n$, $ \mathbf{w_k}$ are Wiener filter coefficients ($ N+1$ of them total, for $ N$ taps filter) and $ \mathbf{x_{n-k}}$ are reference channel values for the time step $ n$ and previous $ N$ steps. Note that at any given $ k$, $ \mathbf{w_k}$ and $ \mathbf{x_{n-k}}$ are  the size of number of reference channels (i.e. they contain coefficients for and measurements in, correspondingly, each reference channel at the time $ k$). Therefore, $ \mathbf{w_k}\cdot \mathbf{x_{n-k}}$ is a scalar product for each $ k$. In other words there's another sum contained in this formula. 

The number of time steps is typically much larger than the number of references (and certainly in our case with 2,160,000 time steps and 3 references), therefore it is more computationally efficient to rearrange this formula in the following way:

$$ \hat{y_n} = \sum_{m=1}^{M} \sum_{k=0}^{N}{w^m_k}\cdot x^{n-k}_m,$$

where $M$ is the number of references. Now the inner dot product (inner sum), contains long vectors of size $ N$ (time steps), which can be efficiently multiplied using \texttt{numpy.dot} function. Such simple rearrangement of sums considerably speeds up the application of the Wiener filter. 

But before a Wiener filter can be applied, the coefficients $w^m_k$ have to be derived first. Derivation requires solving the Wiener-Hopf equations, which are given in~\cite{harms} as follows:

$$\mathbf{C_{xx} \cdot w(:)} =  \mathbf{C_{xy}},$$

where $ \mathbf{C_{xx}}$ is a cross-correlation matrix between the reference channels,  $\mathbf{C_{xy}}$ is a cross-correlation vector of target to reference channels and  $\mathbf{w(:)}$  is the NM-dimensional vector that is obtained by concatenating the M columns of the matrix w, where M is number of references.

Details of the estimation of the cross-correlation matrix $ \mathbf{C_{xx}}$ and cross-correlation vector $\mathbf{C_{xy}}$ are provided in~\cite{ligo_wiener_filter}, and are calculated in the same way here. In~\cite{ligo_wiener_filter} the last step to obtain a solution involves inversion of the cross-correlation matrix:
\begin{equation}
    \label{eq:wiener_invert}
    \mathbf{ w(:)}=  \mathbf{C_{xy}}/\mathbf{C_{xx}}.
\end{equation}

The Wiener filter algorithm is already implemented in \texttt{scipy} and \texttt{MATLAB} packages. However, \texttt{scipy.signal.wiener} is a single-input Wiener filter (that can be applied to high dimensional data). 
The neural network does multiple-input, multiple output (MIMO) conversion of the data. The closest we could do to mimic this (and profit from potential correlations between different DOFs) was to create three multiple-input, single-output Wiener filters. Such a filter is implemented in \texttt{MATLAB} as \texttt{MISO\_FIRWIENER}.
However, as noted in LIGO note~\cite{ligo_wiener_filter}, the calculation in this method is inefficient as it does not exploit Block-Toeplitz matrix structure, described below. For repeated calculation of very large Wiener filters (each with 15,000 taps) it was necessary to improve performance upon that of the calculation referenced. 

In~\cite{ligo_wiener_filter} matrix inversion is performed with standard MATLAB algorithms and we found it to be prohibitively computationally expensive for Wiener filters with thousands of taps, independently of whether done in MATLAB or with other tools. As hinted in~\cite{ligo_wiener_filter}, the calculation can be sped up by exploiting the Block-Toeplitz structure of the matrix $\mathbf{C_{xx}}$. As an illustration, for example with 4 references (4 blocks) the structure looks like this:

$$
\begin{bmatrix}

\begin{bmatrix}
a_0     & a_{-1}  & \cdots  & a_{-(n-1)} \\
a_1     & \ddots  & \ddots  & \vdots \\
\vdots  & \ddots  & \ddots  & a_{-1} \\
a_{n-1} & \cdots  & a_1     & a_0
\end{bmatrix}

&

\begin{bmatrix}
c_0     & c_{-1}  & \cdots  & c_{-(n-1)} \\
c_1     & \ddots  & \ddots  & \vdots \\
\vdots  & \ddots  & \ddots  & c_{-1} \\
c_{n-1} & \cdots  & c_1     & c_0
\end{bmatrix}

\\

\\

\begin{bmatrix}
b_0     & b_{-1}  & \cdots  & b_{-(n-1)} \\
b_1     & \ddots  & \ddots  & \vdots \\
\vdots  & \ddots  & \ddots  & b_{-1} \\
b_{n-1} & \cdots  & b_1     & b_0
\end{bmatrix}

&

\begin{bmatrix}
d_0     & d_{-1}  & \cdots  & d_{-(n-1)} \\
d_1     & \ddots  & \ddots  & \vdots \\
\vdots  & \ddots  & \ddots  & d_{-1} \\
d_{n-1} & \cdots  & d_1     & d_0
\end{bmatrix}

\end{bmatrix}
$$

Each block in $\mathbf{C_{xx}}$ matrix is a Toeplitz matrix, which is fully defined by its first row and first column. This can be exploited to drastically reduce memory requirements for keeping elements of $\mathbf{C_{xx}}$ matrix, because the number of its unique elements is only $M^2\times 2(N+1)$ compared to $M(N+1) \times M(N+1)$ elements for a generic matrix of the same size.

Once the unique $M^2\times 2(N+1)$ elements of the cross-correlation matrix are generated, we used iterative solvers in \texttt{scipy.linalg.sparse} python module to find the solution to the equation~\ref{eq:wiener_invert}. Notably, these iterative solvers do not require knowledge of the entire matrix for each iteration. This is exploited by using \texttt{scipy.sparse.linalg.LinearOperator} functionality to define a linear operator that specifies a rule for the result of the product of the matrix $\mathbf{C_{xx}}$ with an arbitrary vector $\mathbf{v}$. Because of the symmetries of the Block-Toeplitz matrix explained above, such linear operator in combination with unique matrix elements takes up much less memory than the original matrix. The linear operator was implemented as a python function that does the following:
\begin{enumerate}
\item Multiplication is reduced to block-by-block multiplication, which is possible for a matrix of size $M (N+1)\times M(N+1)$ and a vector  $\mathbf{v}$ of size $M (N+1)$;
\item For each block, an efficient multiplication using  \texttt{scipy.linalg.matmul\_toeplitz} is performed, requiring only the unique first column and first row values of the (current) Toeplitz matrix.
\end{enumerate}

Finally with the linear operator defined as above, the equation~\ref{eq:wiener_invert} is solved with \texttt{scipy.sparse.linalg.gmres} solver. Other solvers were tested on ``mock'' data, and \texttt{gmres} was found to be the best-performing (lowest broad-spectrum noise) in different scenarios.

The algorithm described above is implemented in open-source software package \texttt{spicypy}~\cite{spicypy} in \texttt{spicypy.signal.wiener\_filter} module, with example of usage provided in~\cite{spicypy_wf_example}.

\subsection{\label{sec:wiener_filter_best}Choosing the best-performing Wiener filter}

To choose the best-performing Wiener filter, we apply each of the 143 candidates to the total time series of input data, containing 2,160,000 time steps. We then calculate the ASD spectrum of the output data and of the time series produced by Wiener filters using the Logarithmic Power Spectrum Density (LPSD) algorithm~\cite{lpsd}, originally implemented in \texttt{LTPDA} package for LISA mission~\cite{ltpda} and currently also implemented in \texttt{spicypy}~\cite{spicypy}.

Then calculation of the mean squared error (MSE) is done as following:

$$ \text{MSE}_j = \sqrt{ \frac{1}{143}  \sum_{i=0}^{L}  \left[\text{ASD}(\text{output})_i - \text{ASD}(\text{WF}_j(\text{input}))_i \right]^2}, $$

where $i$ is the frequency bin index for the ASD ($L$ bins total) and $j$ is the Wiener filter index (143 total). Using logarithmic binning of the LPSD algorithm here results in a more ``natural'' MSE definition with more sparse binning in high frequency and more frequent in low frequencies (resulting in $\approx$constant binning on a log scale). 

The best filter selection algorithm is also implemented in \texttt{spicypy}~\cite{spicypy}, integrated into Wiener filter generation. By default, if the time series are longer than the number of taps of the filter, multiple Wiener filters will be generated, and the best-performing filter will be found.

\subsection{\label{sec:wiener_filter_references}Effects of number and quality of references on Wiener filter performance}

When more references are added to our Wiener filter implementation, especially low coherence references (such as $X$ and $Y$ ground data for a $Z$-direction Wiener filter), the broad-spectrum noise in the resulting time series increases. Possible reason is due to errors in the correlation estimation; this seems to fit together with the fact that extra noise depends on number of taps and is less pronounced for filters with more taps (and therefore more data for coherence estimation). Detailed analysis of this effect on mock data is provided in \texttt{spicypy} usage example~\cite{spicypy_wf_example}.

\section{\label{sec:nn_wf_comparison}Wiener filter and Neural Network Performance comparison}

 We have used two 6-hour stretches of passive isolation data over two consecutive nights collected in the same measurement campaign as used for plots in Fig.~\ref{fig:passive_iso_1h}. The first 6 hours are used to train a neural network as described in Section~\ref{sec:nn} that takes in 150~s of data (15,000 samples) for 3-axis ground motion at a time and produces 150~s of 3-axis inferred motion on the table. We also construct an equivalent Wiener filter (with 15,000 taps), as described in Section~\ref{sec:wiener_filter}. Out of the other 6 hours of the data, 3 hours were reserved as a validation dataset to tune neural network hyper-parameters (see Section~\ref{sec:nn}). Data from the remaining 3 hours were used in comparison of the results in this section. We used 1 hour of data to compute ASDs and calculate performance metrics of Wiener filters and neural networks, and a random few second stretches of time series for time series plots.

\begin{figure}[hbtp]
\includegraphics[scale=0.4]{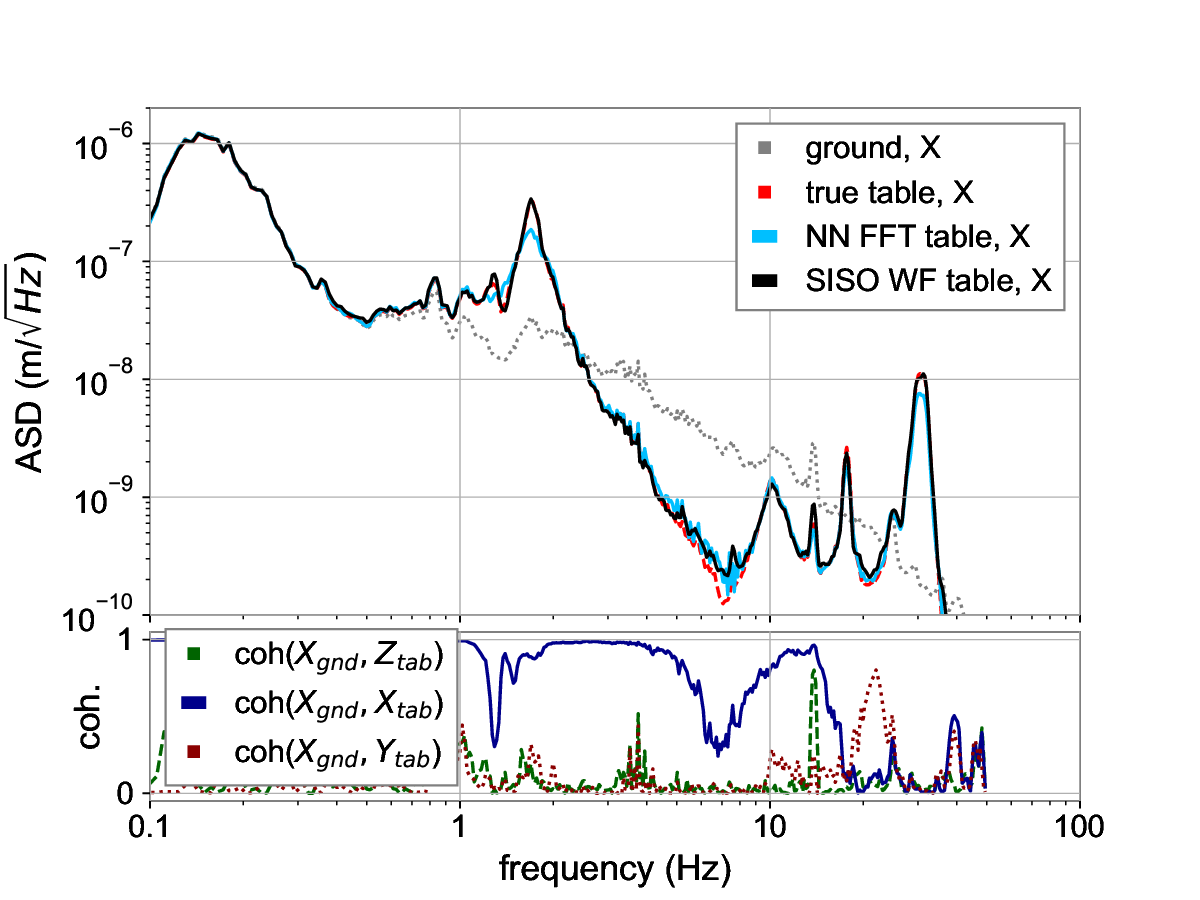}% Here is how to import EPS art
\caption{\label{fig:wf_nn_east}Logarithmic ASD for $X$ axis for the best-performing Wiener filter (black solid line), best-performing neural network (light blue solid line), ground data (gray dotted line) and optical table data (red dashed line). Coherence of the $X$ axis on the ground with all the axes on the table is shown in the subplot (solid line with the $X$ axis, dashed line with the $Z$ axis and dotted line with the $Y$ axis).}
\end{figure}

\begin{figure}[hbtp]
\includegraphics[scale=0.4]{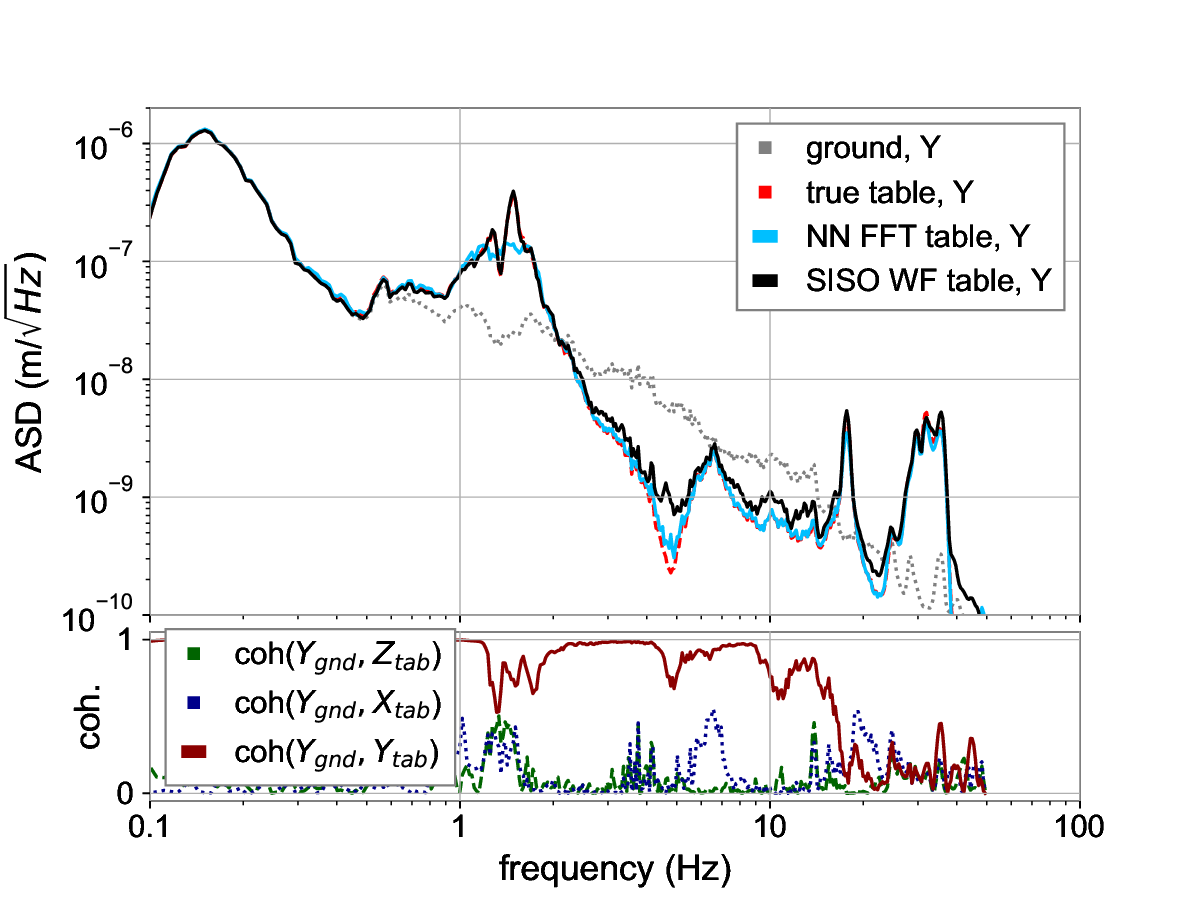}% Here is how to import EPS art
\caption{\label{fig:wf_nn_north} Logarithmic ASD for $Y$ axis for the best-performing Wiener filter (black solid line), best-performing neural network (light blue solid line), ground data (gray dotted line) and optical table data (red dashed line). Coherence of the $Y$ axis on the ground with all the axes on the table is shown in the subplot (solid line with the $Y$ axis, dashed line with the $Z$ axis and dotted line with the $X$ axis).}
\end{figure}

\begin{figure}[hbtp]
\includegraphics[scale=0.4]{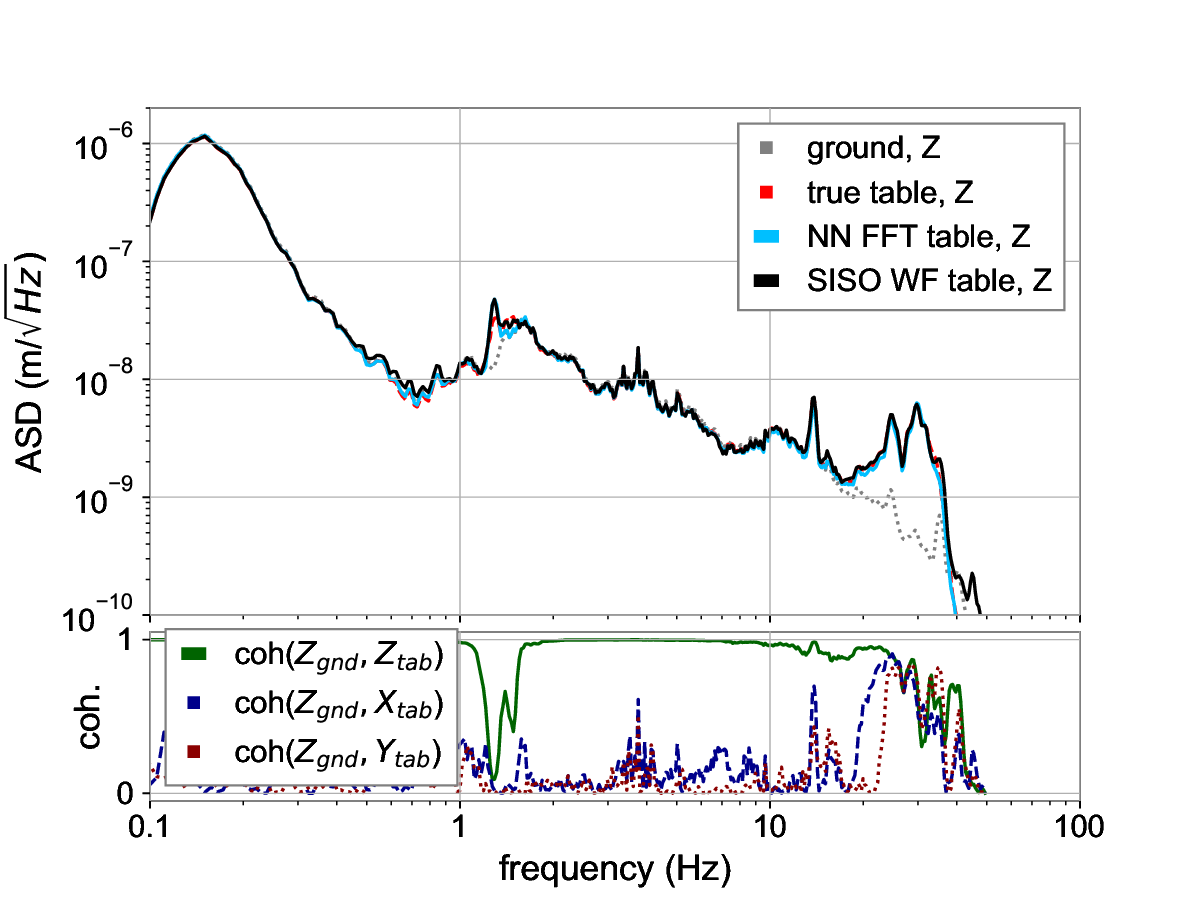}% Here is how to import EPS art
\caption{\label{fig:wf_nn_vert} Logarithmic ASD for $Z$ axis for the best-performing Wiener filter (black solid line), best-performing neural network (light blue solid line), ground data (gray dotted line) and optical table data (red dashed line). Coherence of the $Z$ axis on the ground with all the axes on the table is shown in the subplot (solid line with the $Z$ axis, dashed line with the $X$ axis and dotted line with the $Y$ axis).}
\end{figure}

 All data, originally in units of $\text{digits}/\text{s}$ recorded by the seismometers, were first normalized to improve numerical accuracy. The data were divided by the standard deviation for 6 hours of training data, calculated separately for data on the ground and on the table, because the measurement was performed using different seismometers (STS 2.5 and Trillium 360, respectively). The data were also high-passed at 0.1 Hz to remove noise-dominated lower frequencies, and considering that the first significant feature in the data is the microseism peak at a slightly higher frequency of $\sim0.15$ Hz. Then, after the time series are passed through either the neural network or Wiener filter, the unit of $\text{digits}/\text{s}$ is restored by multiplying by the same standard deviation. After that, the respective seismometer transfer function is applied to finally convert data into $\text{m}/\sqrt{\text{Hz}}$ displacement.

The best-performing neural network turned out to be the one with FFT-based loss function. It is shown in comparison to the best-performing Wiener filter in Fig.~\ref{fig:wf_nn_east}~-~\ref{fig:wf_nn_vert}. 
There's a single neural network that performs the transformation $(X_\text{gnd}, Y_\text{gnd}, Z_\text{gnd}) \rightarrow (X_\text{table}, Y_\text{table}, Z_\text{table})$. 
However, SISO Wiener filters were used for these plots, i.e. 
$ X_\text{gnd} \rightarrow X_\text{table} $, 
$ Y_\text{gnd} \rightarrow Y_\text{table} $, 
$ Z_\text{gnd} \rightarrow Z_\text{table} $, 
because the MISO filters $ (X_\text{gnd}, Y_\text{gnd}, Z_\text{gnd})  \rightarrow X_\text{table} $ (and similar for other DOFs) were found to be severely limited by the broad-spectrum noise described in Section~\ref{sec:wiener_filter_references}. This is due to low coherence of the other reference channels (e.g. $X_\text{gnd}$, $Y_\text{gnd}$ with $Y_\text{table}$), as can be seen in coherence subplot of Fig.~\ref{fig:wf_nn_east}~--~\ref{fig:wf_nn_vert}. 

We can see that for the $Z$ axis, performance of both neural network and Wiener filters is nearly identical and very good, which is not surprising because in that axis there's almost no passive isolation and the motion is directly coupled. For the other axes, we can see that the neural network outperforms Wiener filters for some frequency regions, notably between around 2 and 10 Hz and especially prominent for the $Y$ axis. These regions feature passive isolation anti-resonance with small amplitudes and reduced coherence, and potentially small nonlinearities, which a neural network can apparently handle better. That said, amplitudes of higher peaks tend to be somewhat underestimated by the neural network, and better estimated by Wiener filters.

\begin{figure}[b]
\includegraphics[scale=0.4]{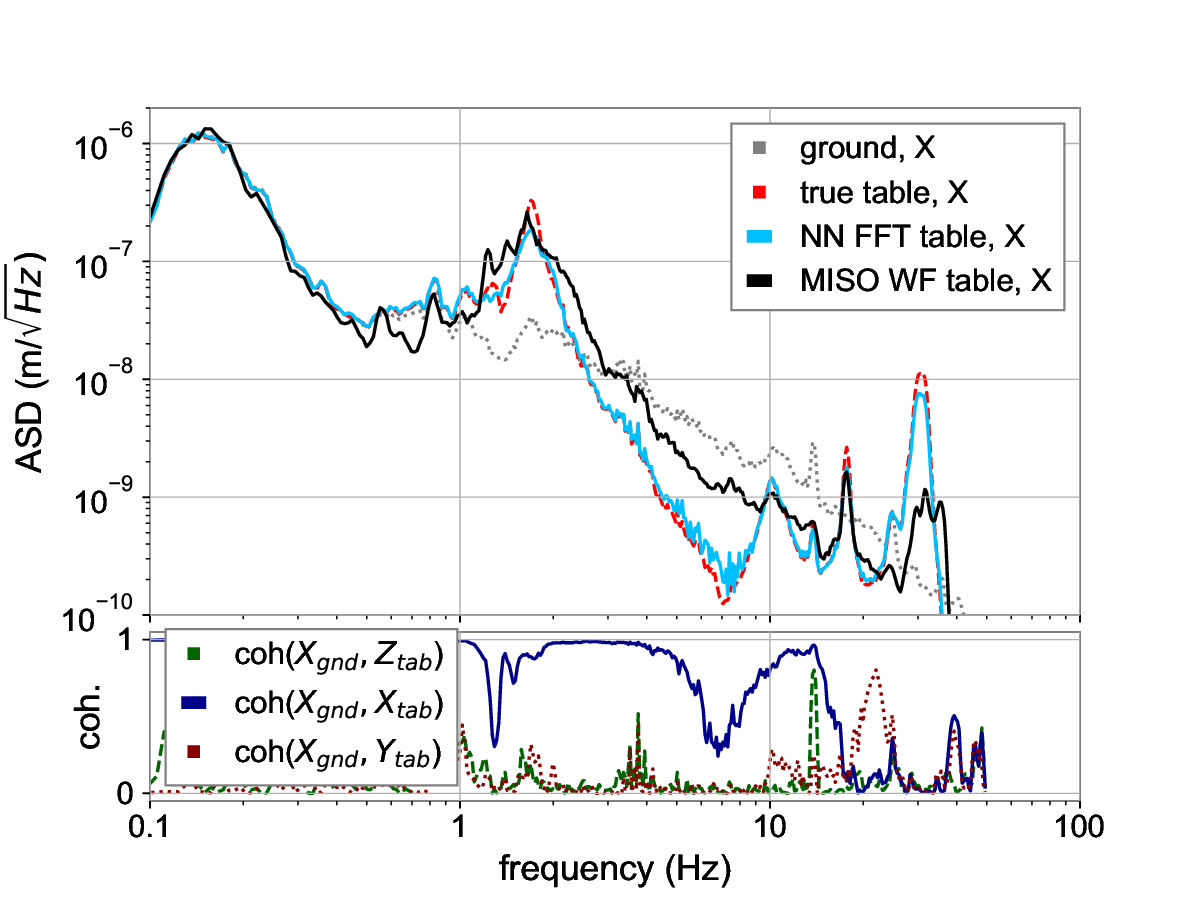}
\caption{\label{fig:wf_nn_east_3ref}Logarithmic ASD for $X$ axis for the best-performing MISO Wiener filter (black solid line), best-performing neural network (light blue solid line), ground data (gray dotted line) and optical table data (red dashed line). Coherence of the $X$ axis on the ground with all the axes on the table is shown in the subplot (solid line with the $X$ axis, dashed line with the $Z$ axis and dotted line with the $Y$ axis).}
\end{figure}

\begin{figure}[b]
\includegraphics[scale=0.4]{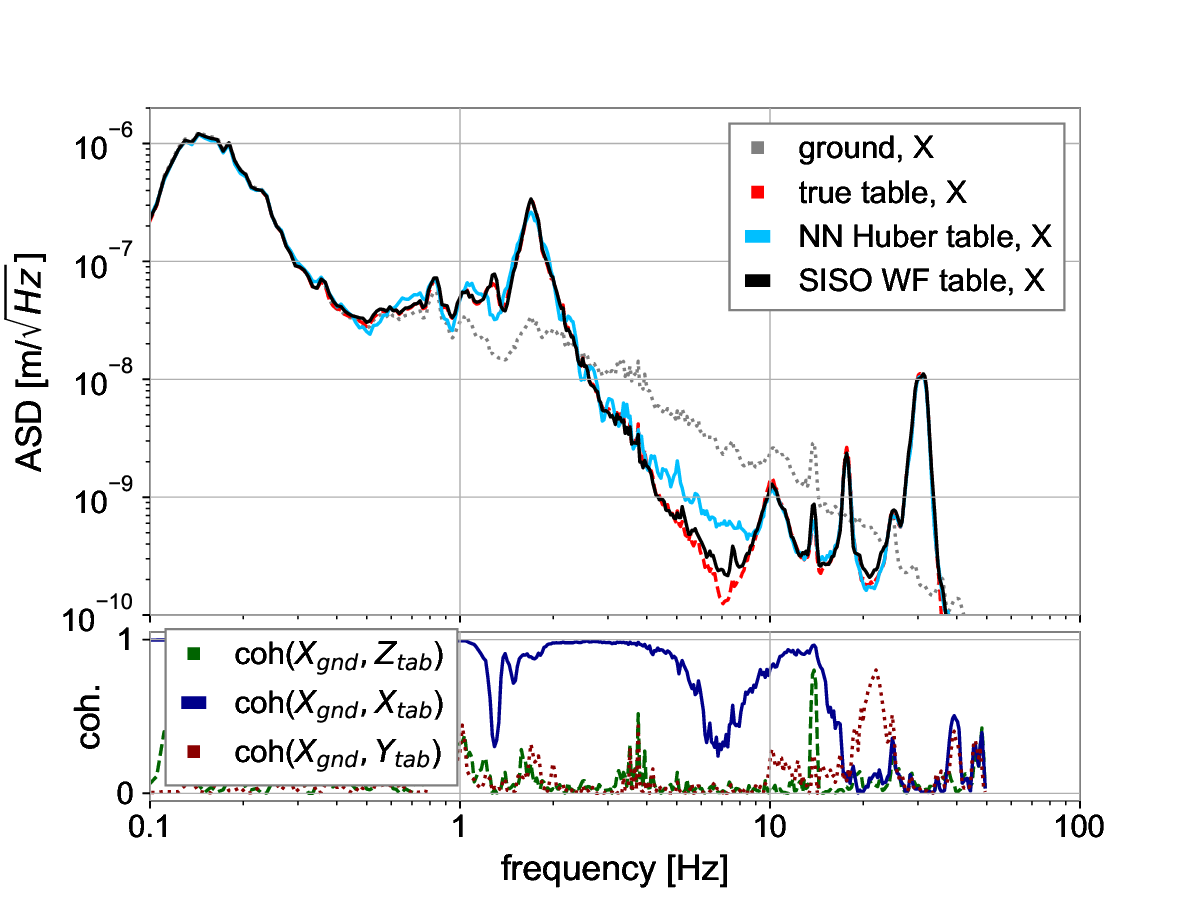}% Here is how to import EPS art
\caption{\label{fig:ASD_of_Huber_loss_x} Logarithmic ASD for $X$ axis for the best-performing Wiener filter (black solid line), CNN with Huber loss (light blue solid line), ground data (gray dotted line) and optical table data (red dashed line). Coherence of the $X$ axis on the ground with all the axes on the table is shown in the subplot. (solid line with the $X$ axis, dashed line with the $Z$ axis and dotted line with the $Y$ axis).}
\end{figure}

There's an interesting small feature (peak) for $X$ axis (in Fig.~\ref{fig:wf_nn_east}) between 10 and 20 Hz, where we can see a coupling from $Z$ direction from coherence subplot. The neural network has information about the $Z$ axis (unlike a SISO WF) and performs better here. The extra noise from additional references for the Wiener filter unfortunately dominates all expected potential gains from additional information in other axes. Besides physical couplings, another source for such information could be due to seismometer's true axes $U,V,W$ not coinciding with $X,Y,Z$. Readings in $X,Y,Z$ are actually obtained via a coordinate transformation of the data, which leaves a possibility of coupling due to imperfect transformation. 

In Fig.~\ref{fig:wf_nn_east_3ref} we show an example of $(X_\text{gnd}, Y_\text{gnd}, Z_\text{gnd})  \rightarrow X_\text{table}$ MISO filter in comparison with the same as above neural network. Time series produced by this MISO filter have been normalized by applying a constant factor, minimizing spectral residuals on training data. Without normalization, the result is off by orders of magnitude. Even with normalization, it can be seen that its performance is still significantly worse than a simple SISO filter shown in Fig.~\ref{fig:wf_nn_east}, where no such extra normalization factors were needed. 

It should be noted that the algorithm producing MISO Wiener filter has been tested in more favorable conditions where all references are correlated to the output and produces a good result (see \texttt{spicypy} usage example~\cite{spicypy_wf_example}), which suggests that the issue here lies with the data, namely the two noisy references, $Y_\text{gnd}$ and $Z_\text{gnd}$, with very small correlation to the output, $X_\text{table}$.

The ASD result for the network with the Huber loss (described in Section~\ref{sec:nn}) is shown for the $X$ axis in Fig.~\ref{fig:ASD_of_Huber_loss_x}, just to pick one as an example. While the peak amplitudes are inferred somewhat better than for the network with FFT-based loss, it is visible that overall this network shows inferior performance to both Wiener filters and the network trained on FFT-based loss.

\begin{figure}[h]
\includegraphics[scale=0.18]{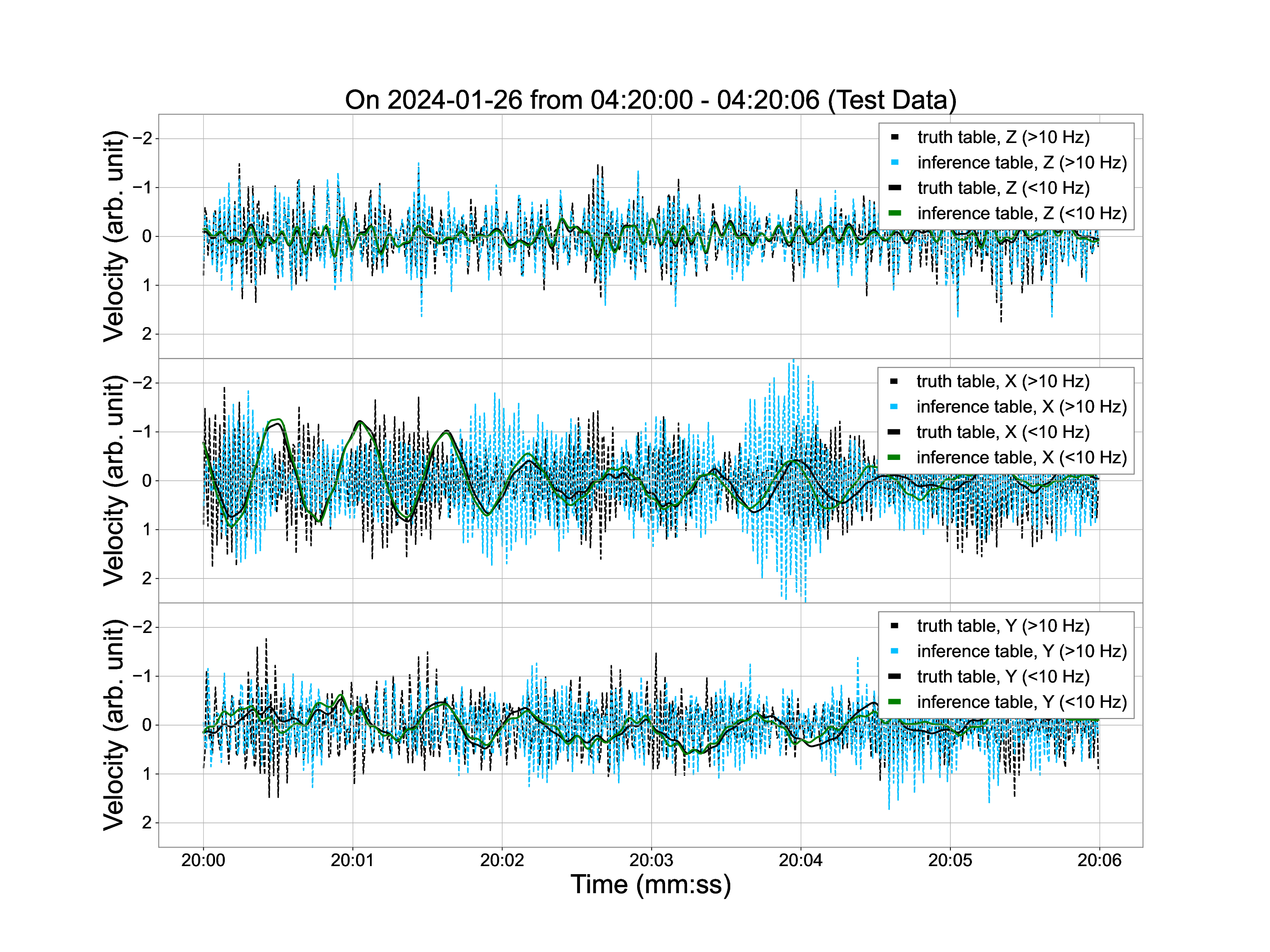}% Here is how to import EPS art
\caption{\label{fig:TS_of_WF} A six seconds time series showing the inference by the SISO Wiener filter of the table motion from ground motion and the true table motion. Each shown as two traces: low-frequency (low-passed at 10 Hz, solid lines) and high-frequency (high-passed at 10 Hz, dashed lines).}
\end{figure}

\begin{figure}[h]
\includegraphics[scale=0.18]{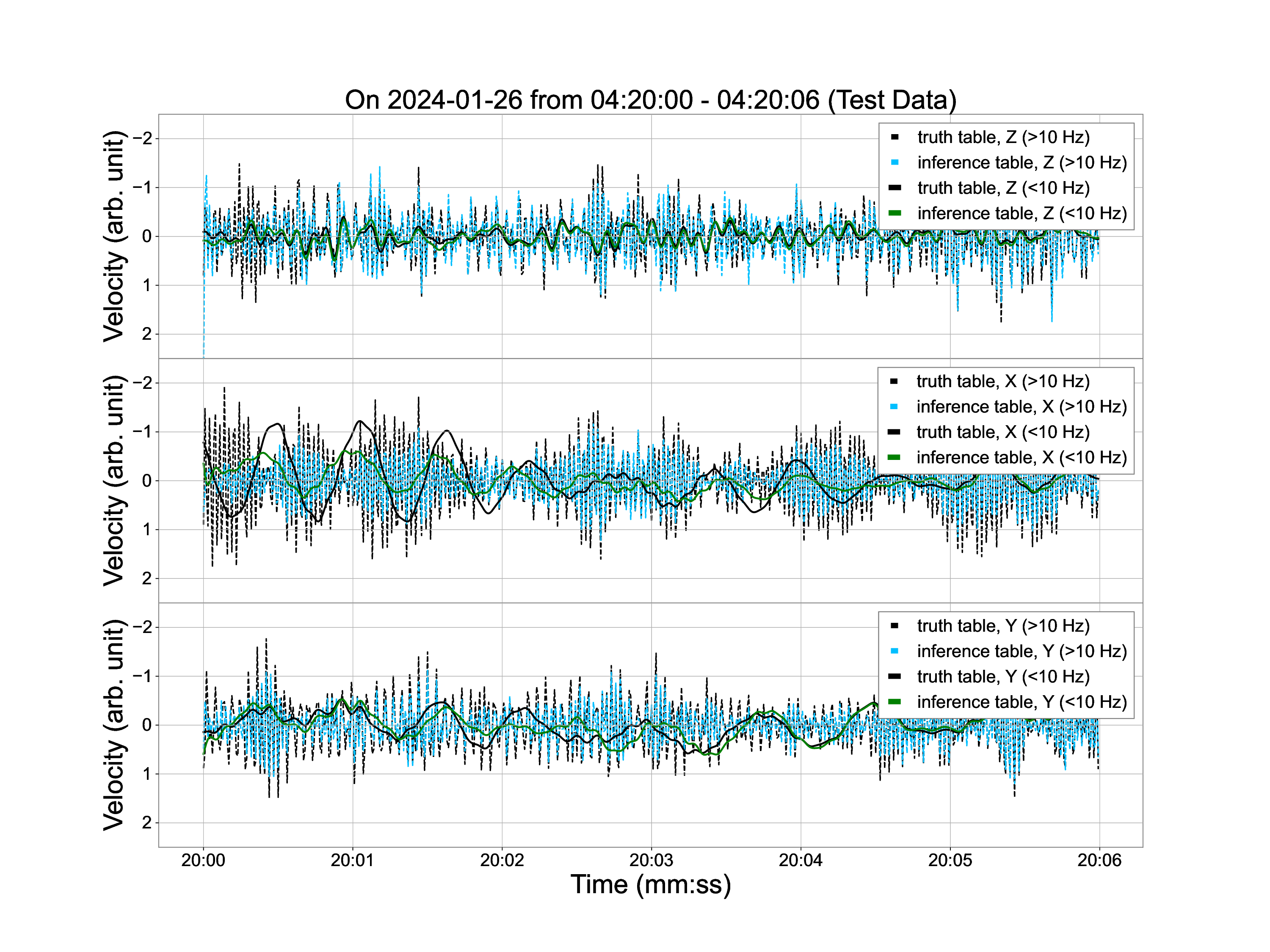}% Here is how to import EPS art
\caption{\label{fig:TS_of_FFT_loss} A six seconds time series showing the inference motion by the CNN trained with the FFT loss of the table motion from ground motion and the true table motion. Each shown as two traces: low-frequency (low-passed at 10 Hz, solid lines) and high-frequency (high-passed at 10 Hz, dashed lines).}
\end{figure}

\begin{figure}[h]
\includegraphics[scale=0.18]{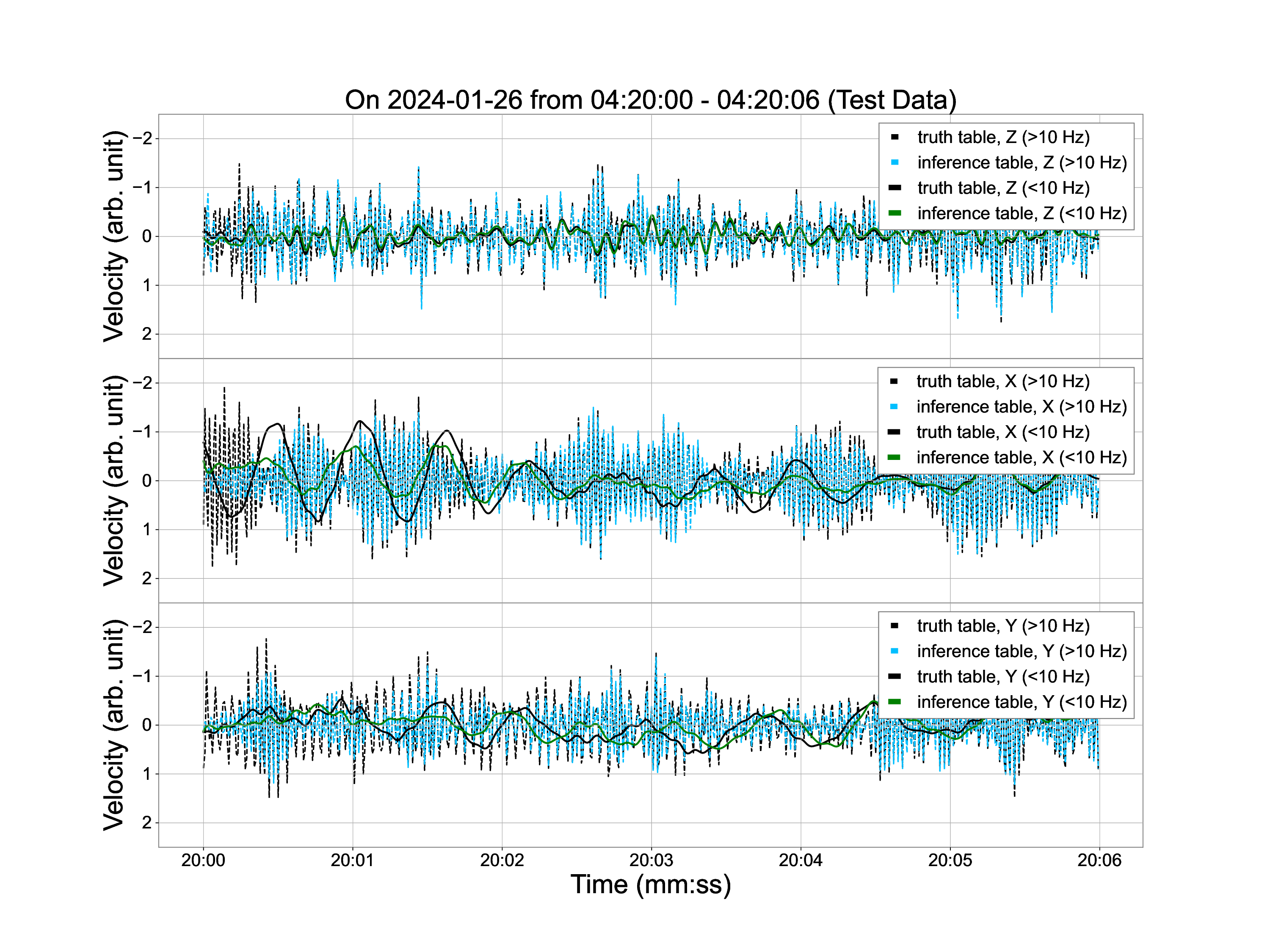}% Here is how to import EPS art
\caption{\label{fig:TS_of_Huber_loss_x} A six seconds time series showing the inference motion by the CNN trained with the Huber loss of the table motion from ground motion and the true table motion. Each shown as two traces: low-frequency (low-passed at 10 Hz, solid lines) and high-frequency (high-passed at 10 Hz, dashed lines).}
\end{figure}

Time series performance for the SISO Wiener filters is shown in Fig.~\ref{fig:TS_of_WF},  and  for the neural networks in Fig.~\ref{fig:TS_of_FFT_loss} (FFT loss) and in Fig.~\ref{fig:TS_of_Huber_loss_x} (Huber loss). To aid the comparison time series are presented as two traces: low-frequency (low-passed at 10 Hz) and high-frequency (high-passed at 10 Hz). It is visible that Huber-loss-trained network has slightly worse performance than other algorithms for time series. The Wiener filter shows very good performance in low-frequency ($<10$ Hz trace), better than neural networks, especially for the horizontal axes. We can see that the FFT-loss-trained network shows better time series performance than the Huber-loss-trained, but does not reproduce the amplitude of an approximately 2~Hz oscillation for the $X$-axis in the low-frequency trace. This is consistent with the underestimation of peaks around 2~Hz seen in ASD. However for higher frequencies it is difficult to make the conclusion. On these plots one can see that high-frequency time series of Wiener filter have somewhat larger spread than that of the FFT-loss-trained network, especially visible for the $X$ axis. 

The inference time with neural network is significantly shorter than with Wiener filter on the same data in the setup that we have used. Table~\ref{table:comparisonCNNWF} shows performance comparison with loss values and typical timings for the same 1 hour of data as used for the ASD plots. Timings cannot be compared directly however, because the neural network inference ran on a GPU (NVIDIA GeForce RTX 3090), while the Wiener filter inference was performed on a CPU (Intel Core i7-9800X, 3.80~GHz).

\begin{table}[hbtp]
    \centering
    \begin{tabular}{l||c|c|c}
             & CNN FFT loss & CNN Huber loss & Wiener filter \\ \hline \hline
    MAE      & 0.2769         & 0.1729           &    0.6874       \\ \hline
    MSE      & 0.1261         & 0.0492           &    0.9433       \\ \hline
    RMSE     & 0.3551         & 0.2217           &    0.9713     \\ \hline
    FFT loss & $0.5750 \cdot 10^{-3}$         & $0.9880 \cdot 10^{-3}$       &    $0.7568  \cdot 10^{-3}$       \\ \hline
    Huber loss & 0.0162       & 0.0094           &    0.0432       \\ \hline
    Run time & $0.34 \pm 0.01$ s & $0.34 \pm 0.01$ s & $24.01 \pm 0.61$ s    
    \end{tabular}
    \caption{Performance comparison of the CNN with different loss functions and Wiener filter shown in metrics calculated with their output and the true data (same 1 hour as used for the ASD plots), the different losses and a run time performance.}\label{table:comparisonCNNWF}
\end{table}

\section{\label{sec:improvements}Conclusion and outlook}

In this paper we investigated seismic motion propagation through a passively isolated mechanical system, using Wiener filters and convolutional neural networks. The mechnical system studied is a gravitational-wave detector technology testbed ``VATIGrav'' that is currently being commissioned at University of Hamburg. VATIGrav consists of a seismically-isolated vacuum chamber and an optical table inside. It will be used as testbed for displacement and inertial sensors, as well as control algorithms, to improve the low-frequency performance of current and future ground-based gravitational wave detectors.e studied how well we could infer seismic motion on the table from the motion on the ground. We used 150 seconds of ground motion data sampled at 100 Hz to infer 150 seconds of motion on the table at the same sampling rate with both approaches, comparing the performance in the ASD plots for 1 hour of data and randomly chosen few seconds of time series. 

Our goal was to achieve good performance in broad frequency range, spanning from 0.1 to 50 Hz. We have used a CNN with time-dilation layers to strengthen the performance for lower frequencies, and tried two different loss functions. The first one, Huber loss, is calculating the difference in amplitude for each time sample, with quadratic response for small differences and linear response for large differences, to ignore outliers. The second is a custom loss function that is calculating the difference of the FFTs in amplitude and phase. All neural networks take 3-axis motion data as input and output 3-axis motion data.

Broad frequency range and relatively long time series presented a challenge for deriving Wiener filters, the algorithm had to be optimized using block-Toeplitz structure of the Wiener-Hopf equations to be computationally efficient. We compute SISO Wiener filters (one axis of ground motion as input and the same axis as output) and MISO (all axes as input and one axis as output).

We found that the MISO Wiener filters perform quite poorly for our data, where the coherence between different channels (motion axes) is low. Adding extra reference axes introduces penalty in extra noise which in this case is larger than any potential gains from information about cross-couplings. SISO Wiener filters, on the other hand, performed quite well both as seen on ASD plots and in time series directly.

Among neural networks, the Huber-loss-trained network showed good but somewhat inferior performance. The FFT-loss-based network was significantly better and outperformed the Wiener filters in some frequency ranges, as seen in the ASD plots. However it tended to slightly underestimate the peaks, especially in lower frequencies, and hence low-frequency time series plots for Wiener filters show better agreement. Overall we find that the FFT-loss-based neural network showed advantage over the Wiener filter for low-amplitude low-coherence regions, but does not surpass its performance everywhere. It is in principle possible that with further optimization the performance of the neural network could be further improved. It is also worth considering other network architectures, such as encoder-decoder approaches, e.g. the U-net architecture, which may be well-suited for data with input and output of the same dimensionality~\cite{UNet}. Another angle not explored here is, in addition to inference, also forecasting time series a few seconds in the future, which can be useful for control. Finally there are alternative approaches to Wiener filters and neural networks to study a mechanical system using measurement data. They include simple linear algorithms, for example least mean squares (LMS) algorithms~\cite{LMS_filters} (a class of adaptive filters), as well as more general non-linear algorithms, such as Sparse identification of nonlinear dynamics (SINDy)~\cite{SINDy}.

\section{\label{sec:ack}Acknowledgements}

 This research was funded by the Deutsche Forschungsgemeinschaft (DFG, German Research Foundation) under Germany's Excellence Strategy---EXC 2121 ``Quantum Universe''---390833306, and by the German Federal Ministry of Education and Research (BMBF, Projects 05A20GU5 and 05A23GU5). VATIGrav is a testbed facility for technology development for current and future gravitational wave detectors, funded by the University of Hamburg/State of Hamburg and the FG, project number 455096128.

 In this work we benefited from computing resources of Physnet Computing Cluster~\cite{physnet_cluster}, specifically in training CNNs on GPUs. We also benefited from wandb~\cite{wandb} online training monitoring system and hyper-parameter optimization (sweep).

 We thank Jana Klinge and Alexander Bauer from Department of Earth System Sciences of University of Hamburg, Gabriele Vajente (Caltech), and Terrence Tsang (Cardiff University) for fruitful discussions about neural network architectures and loss functions for time series. We also thank Rana Adhikari (Caltech) for useful insights into the workings of STACIS active isolation system.

\bibliography{paper}% Produces the bibliography via BibTeX.

\end{document}